# Sputtered Spontaneously Nano-porous VO$_2$-based Films *via* PTFE Self-Template: Localized Surface Plasmon Resonance Induced Robust Optical Performance for Solar Glazing Application

Shiwei Long, [a,b,c] Xun Cao, [a,b,*] Rong Huang,[d] Fang Xu, [a,b,c] Ning Li,[e] Aibin Huang, [a,b] Guangyao Sun, [a,b] Shanhu Bao, [a,b] Hongjie Luo[f] and Ping Jin [a,b,g,*]

[a] State Key Laboratory of High Performance Ceramics and Superfine Microstructure, Shanghai institute of Ceramics, Chinese Academy of Sciences, Shanghai, 200050, China.
[b] Research Center for Industrial Ceramics, Shanghai Institute of Ceramics, Chinese Academy of Sciences, Shanghai 200050, China
[c] University of Chinese Academy of Sciences, Beijing 100049, China
[d] Key Laboratory of Polar Materials and Devices, Ministry of Education, East China Normal University, Shanghai 200241, China
[e] Department of Materials Science and Engineering, College of science, China University of Petroleum Beijing, No. 18 Fuxue RD, Beijing 102249, China
[f] School of Materials Science and Engineering, Shanghai University, Shangda Road. 99, Baoshan, Shanghai 200444, China
[g] Materials Research Institute for Sustainable Development, National Institute of Advanced Industrial Science and Technology, Nagoya 463-8560, Japan

**Corresponding authors**
State Key Laboratory of High Performance Ceramics and Superfine Microstructure, Shanghai Institute of Ceramics, Chinese Academy of Sciences, Shanghai 200050, China
*E-mail：cxun@mail.sic.ac.cn (X. Cao)
*E-mail：p-jin@mail.sic.ac.cn (P. Jin)

**Abstract**

The PTFE (Teflon) has been selected as the self-template structural material in preparation of $VO_2$ films using reactive magnetron sputtering systems and post annealing progress. Spontaneous random nano-porous structures of $VO_2$ films growing on quartz glasses have been deliberately established *via* bottom-up processing through this novel and facile approach. The nano-porous $VO_2$ films exhibit an excellent optical performance based on localized surface plasmon resonance (LSPR), with ultrahigh luminous transmittance ($T_{lum}$) up to 78.0% and the promoted solar modulation ability ($\Delta T_{sol}$) of 14.1%. Meanwhile, the ingenious microstructure of film provides an antireflection function from multiple perspectives in visible light, with the potential of the windshield on vehicles for smart solar modulation. The nano-porous films expand the practical application of thermochromic $VO_2$ to a fire-new field, breaking the optical performance envelope of single-layer dense $VO_2$ film away, and offering a universal method to prepare homogeneous nano-porous structures for thin films.

## 1 Introduction

Intelligent windows, or smart windows, can automatically manage the heat by modulating the amount of solar radiation (especially the near-infrared region) in response to environmental temperature variations, exhibit potential application in building energy-saving and temperature control.[1-3] Among diverse thermochromic materials, vanadium dioxide ($VO_2$) was regarded as the representative candidate due to the reversible phase change from an infrared-transmitted semiconductor-state ($VO_2$(M)) to infrared-reflective metal-state ($VO_2$(R)) when below/over the critical transition temperature (~340K), while nearly maintains the transmittance in luminous range.[4,5] Therefore, $VO_2$ films can be capable to impede unnecessary heat at high temperature while allow the incident solar radiation into indoors automatically, displays an ideal performance for energy savings.[6,7] However, the relatively high phase transition temperature should be decreased to meet the requirement.

Fortunately, bringing in high valence dopants like tungsten can provide an efficiently approach to reduce intrinsic transition temperature (~68°C) of $VO_2$ by modifying the electronic density of states.[8,9] Nevertheless, some other drawbacks, such as the low luminous transmittance ($T_{lum}$) with unsatisfied solar modulation ability ($\Delta T_{sol}$) as well as the native dark brown color of films have still obstructed the actual use of $VO_2$-based smart windows.[10,11] Here, the low luminous transmittance and the native dark brown color of films were probably caused by small optical band gap of $VO_2$, which gives rise to absorption in short-wavelength range.[12,13] A large number of methods have been employed to enhance the optical performance ($T_{lum}$ and $\Delta T_{sol}$) of $VO_2$. Doping is also used here, dopants like Mg or F can increase $T_{lum}$ but make a deterioration of $\Delta T_{sol}$.[14-16] In addition, antireflection layers such as $TiO_2$, $Cr_2O_3$, $WO_3$, have been introduced to enhance both $T_{lum}$ (40-55%) and $\Delta T_{sol}$ simultaneously based on interference effect,[17-20] but the improvement of $T_{lum}$ cannot completely satisfy the requirement, which is at least 60% to avoid extra energy consumption on indoor lighting.[21-23] Therefore, all aforementioned methods have hardly equilibrated the optical performance for both of $T_{lum}$ and $\Delta T_{sol}$.

Surface modification seems one of the most feasible routes to take account of both $T_{lum}$ and $\Delta T_{sol}$. Xie et al.[24] masterly utilized monolayer PS (polystyrene) spheres as template to fabricated periodic porous $VO_2$ films, successfully enhanced the $T_{lum}$ (70.2%), and obtained decent $\Delta T_{sol}$ (7.9%). While Jiang et al.[25] used a dual-phase transformation method to fabricate $VO_2$ films with ordered honeycomb microstructures, achieving the excellent transmittance of 94.5% at 700nm with $\Delta T_{sol}$ of 5.5%. In addition, except ordered surficial structures, some other previous works had also confirmed that random nano-porous structures on $VO_2$ thin film could significantly improve $T_{lum}$ and maintain the $\Delta T_{sol}$, which may largely depend on the interference effect derived from reducing optical constants.[26,27] Thus, nano-porous patterns play a positive role in optimizing the optical properties, especially in $T_{lum}$. For preparing ordered surface morphology, it is necessary to apply several advanced technologies, such as colloidal lithography,[28,29] photolithography[30,31] or ion etching,[32] which may limited by the specific template or size and requirements. In consideration of the high costs and complex processes in fabrication of ordered structures, spontaneous self-template seems a feasible method in actual use, which could form random pores. Besides, $VO_2$ particles and films fabricated by conventional chemical method exhibited a weak adhesion to substrate, which is not suitable to long-term service and shown an inferior durability. Hence, a self-template pattern $VO_2$-based films fabricated by magnetron sputtering may be a wise choice in large-scale preparation.

Inspired by these interesting design concept, we first prepared a spontaneous nano-porous $VO_2$

thin film *via* magnetron sputtering method. Herein, the polytetrafluoroethylene (PTFE, Teflon) was deliberately introduced as the self-template on deposited composite films in consideration of its great chemical stability and non-adhesive properties,[33, 34] which could serve as the pore-forming material on post-annealing process. As is expected, the obtained uniform porous $VO_2$ exhibits ultrahigh luminous transmittance ($T_{lum}$ =78.0% in maximum) and brilliant solar modulation ability ($\Delta T_{sol}$ =14.1%). This ultrahigh $T_{lum}$ has satisfied the requirement of windshield on vehicles, which need to be over 70% to guarantee the safety. In addition, it is exciting that both $T_{lum}$ and $\Delta T_{sol}$ had largely enhanced when design the "double-layer" nano-porous $VO_2$ films structure, which reached excellent $T_{lum}$ of 63.3% and $\Delta T_{sol}$ of 20.1% respectively. We speculate that the Located Surface Plasmon Resonance (LSPR) mainly contribute to these optimizations for optical performance. We believe that the proposed universal approach will further impel the development of practical smart window and intelligent windshield for promotion of energy savings.

## 2 Experimental section

### 2.1 Preparation of Films

$VO_2$ Films were fabricated on 10×10 mm$^2$ quartz glass substrates *via* reactive magnetron sputtering system and the subsequent annealing. $VO_2$ (99.99%) and PTFE (99.95%) targets (diameter of 2 inch) were used for co-deposition while deposition course was carried out using integrated lock-load system. An initial pump-down process was executed to reach the original pressure of $5.0\times10^{-4}$ Pa for deposition chamber. Then, Ar (99.99% pure) gas as well as the Ar (97%) and $O_2$ (3%) mixed gases (99.99% pure) with fixed Ar (Ar+$O_2$)/Ar proportion were introduced to the atmosphere. Co-sputtering then took place at the DC and RF powers of 80W and 60W corresponding to $VO_2$ and PTFE targets. The (Ar+$O_2$)/Ar ratio was kept at 0.6 ($O_2$~1.1%) and the pressure of chamber was kept at 0.6 Pa when the total gas flow is 40 sccm. Co-sputtering was executed at room temperature through entire deposition process for $VO_x$/PTFE mixed films. Post annealing was carried out after deposition in two steps. First, the composite film was annealing at 350°C for 3min, with the heating rate of ~1.5°C /s. Then, annealing temperature rose to 450°C and kept for 5min. Finally, the samples decline to room temperature by furnace cooling. Moreover, when prepared continuous films, the steps were the same with fabricating nano-porous films, but the co-sputtered PTFE was not needed.

### 2.2 Calculation of Optical Properties

The integral value of solar transmittances (350-2600nm) of films and the luminous transmittances ($T_{lum}$, 380-780 nm) were obtained by the formulas

$$T_{sol}= \int \Phi_{sol}(\lambda)T(\lambda)d\lambda / \int \Phi_{sol}(\lambda)d\lambda \qquad (1)$$

$$T_{lum}= \int \Phi_{lum}(\lambda)T(\lambda)d\lambda / \int \Phi_{lum}(\lambda)d\lambda \qquad (2)$$

where $T(\lambda)$ means spectral transmittance at wavelength $\lambda$. $\Phi_{sol}(\lambda)$ is the solar irradiance spectrum for an air mass of 1.5 corresponding to the sun standing 37°above the horizon, and $\Phi_{lum}(\lambda)$ is the standard efficiency function for photopic vison. Herein, the solar modulation ability ($\Delta T_{sol}$) of films can be expressed by

$$\Delta T_{sol} = T_{sol-L} - T_{sol-H} \qquad (3)$$

Where L and H stand for Low and High temperature. In consideration of the room temperature, critical transition temperature (~68℃) as well as the pure phases of the $VO_2$ films, we select 20°C and 90°C as the low/high temperature to maintain the insulator/metal states of the samples in the measurements, respectively.

**2.3 Characterization**

X-ray diffraction (XRD) measurements were conducted on a Rigaku Ultima IV diffractometer with 2θ grazing angle mode using Cu Kα radiation (λ= 0.15418 nm), while the Raman shifts were performed using Raman spectrometer (HORIBA, Lab RAM HR Evolution). Surface topography and root-mean-square (RMS) roughness of films were measured by atomic force microscopy (AFM, SII Nano Technology Ltd, Nanonavi Π) apparatus in tapping mode. To confirm the thickness and determine the microstructure of the films, cross-section images and surface morphology were observed using field-emission scanning electron microscopy (SEM, Hitachi SU8220 and FEI, Helios G4 UX). Transmittances spectra (350-2600 nm) and hysteresis loops (2000nm) of the as-deposited films were obtained using a UV-vis-NIR spectrophotometer (Hitachi Corp., model UV-4100) equipped with an accessory heater. The temperature was measured precisely with a temperature sensor in contact with the surface of films, commanded by a temperature controlling unit. Angle-dependent transmittances in visible light were measured by Angle-resolved spectrum measurement system, as described in other literatures.[78-80]

**2.4 Simulation of Spectra**

The optical constants ($n, k$) of dense $VO_2$ (M/R) film were taken from our previous deposited film (**Figure S11**) from 350 to 2500 nm. The simulated spectra were calculated by the optical constants of void film and $VO_2$ particle film. And the transmittance for **model 3** is assumed to be 100% over all wavelength range (350-2500nm). The final model was combining the former three models together by synthesizing their spectra with different percentages.

## 3 Results and Discussion

**3.1 Morphology and Structure of Film**

The field emission scanning electron micrograph (FESEM) image in **Figure 1a** shows the large-scale surface morphology and structure of the prepared $VO_2$ film. One can see that random pores distribute on the film homogeneously. The amplified image of red rectangle area in **Figure 1a** was illustrated in **Figure 1b**, manifests that the random wormlike pores are on nanoscale with the size of 50nm-200nm (**Figure S1b-d, Supporting Information**), which lead to the formation of plentiful continuous particles. Furthermore, **Figure 1c** and **Figure S1a**, provide the cross-section FESEM image of the film, demonstrating that the surficial particles are teeth-like and a portion of the pores were perforative, with the depth of ~210 nm (film thickness, **Figure S1a**), while some of the pores were not, means that the particles were nearly connected to each other. Atomic force micrograph (AFM) image (**Figure 1c**) could present the surface morphology of the film. The Root-Mean-Square (RMS) roughness is around 15.5 nm, implying that smooth surface was obtained in prepared film. In addition, AFM image also conforms the existence of some pores, to the depth of ~70 nm in the film (**Figure S2b**), in agreement with the FESEM images. **Figure 1d** shows the X-ray diffraction (XRD) pattern of deposited film, and the main peaks can be indexed with (011), (-211) and (220) of $VO_2(M)$ (JCPDS card NO. 72-0514, P21/c, a=0.574 nm, b=0.452 nm, c=0.538 nm, and α=γ=90°, β=122.61°) ,while other peaks can be also indexed to pure $VO_2(M)$. And the XRD pattern is similar to other prepared dense $VO_2$ films (**Figure S2a**). To further conform the structure, Raman shift were executed, displayed in **Figure 1d**. As is expected, almost all known Raman vibration modes of monoclinic $VO_2$ have been observed (144, 194, 226, 262, 310, 342, 389, 445, 500, and 615 $cm^{-1}$).[35, 36] In addition, Raman shift of prepared dense film are also matched well with monoclinic $VO_2$ (**Figure S3**).

## 3.2 Preparation and forming mechanism of porous films

To expound the fabrication process and formation mechanism of this nano-porous film, **Figure 2a** and **Figure 2b** illustrate the preparation method and the speculation of mechanism for pores in post annealing in detail respectively. First of all, quartz glass substrate was rinsed to make it non-pollution and dryness. Then the co-sputtering was carried out using $VO_2$ ceramic and PTFE targets, resulting in depositing a continuous mixed film on quartz glass. Finally, the deliberate post annealing was executed on extraordinary parameters, leading to the formation of $VO_2$ film with random surficial nano-pores and nano-particles, even the perforative pores. Notably, post annealing route is one of the most vital steps. To our knowledge, the PTFE has the lowest adhesive energy in solid states. [33, 37-39] Thus, when continuous $VO_x$ and PTFE mixed film was formed, PTFE may form as nearly nano-scale spherical particles and uniformly distribute in amorphous $VO_x$ film on co-sputtering step, so as to guarantee the lowest surface energy on the interface with $VO_x$,[40, 41] not joint $VO_x$, and form a dispersive system in mixed dense film. Herein, post annealing process was intentionally executed in two steps. Primarily, annealing temperature was slowly risen to ~350°C with the heating rate of ~1.5°C /s, and kept for 3 min. Above the melting point of ~327°C,[42] PTFE would fuse and undergoes endothermic process, and we speculate that PTFE particles may start to melt and combine together adequately to aggregate the lager fusing particles, because this approach would decline the interfacial area and reduce the surface energy to make PTFE reach a metastable state. Then, annealing temperature was elevated to 450°C and maintain 5 min, which is above the sintering point of ~380°C and boiling point of ~400°C for PTFE.[43-45] In this case, PTFE particles may volatilize and ablate in the furnace, constructing the pores and particles from bottom to the surface of films. Further annealing is aim at making the rest of the $VO_x$ transform into crystalline $VO_2$ in the end. In comparison, when we take the primarily annealing step out, the $VO_2$ particles of film collapsed (**Figure S4**), forming a broken surface for film which is full of cracks heterogeneous particles, and we speculate that the improper annealing temperature may be the main cause (**Figure S5**). Therefore, annealing temperature and steps is the crucial point in form of nano-pores and nanoparticles.

## 3.3 Optical Performance

As expected, this kind of porous structure serves as a positive role for optical performance. **Figure 3a** exhibits the transmittance spectra at 20°C and 90°C for prepared nano-porous $VO_2$ film. Here, the film has an ultrahigh transmittance at visible range whether in semiconductor state nor in metallic state. The film has reached an ultrahigh luminous transmittance ($T_{lum}$) of 78.0% ($T_{lum-L}$) and 72.8% ($T_{lum-H}$) at 20 and 90°C respectively, and delivers an even higher solar modulation ability ($\Delta T_{sol}$) of 14.1%, exhibits a splendid optical performance compared with dense $VO_2$ films (**Figure S6a** and **Figure S6b**). Here, we speculate that the formation of Localized Surface Plasmon Resonance (LSPR) absorption peak in metallic states might contribute more to the enhancement of solar modulation efficient. Furthermore, double-layer nano-porous $VO_2$ films (labeled as 2VP, **Figure S6c**) were also designed. The spectra (**Figure 3d**) of 2VP display a larger difference in low/high temperature, especially in near-infrared range, with an obvious LSPR absorption peak. In addition, the coloring areas represent the normalized values of spectral irradiance corresponding to the visible (cyan) and NIR (red) ranges of solar spectra, as well as the eye sensitivity function (yellow) respectively.[11, 46, 47] Under the circumstance, the double-layer nano-porous film shows a $T_{lum-L}$ of 63.3% with a $\Delta T_{sol}$ of 20.1%, exhibits a large enhancement of optical performance in equilibrating luminous transmittance and solar modulation ability. The optical performance of all the prepared films were calculated and collected in **Table 1**, and the nano-porous $VO_2$ film is undisputedly the most excellent films in overall performance. Moreover, the optical energy band gap ($E_g$) based on transmittance spectra had also demonstrate that the $E_g$ values of nano-porous film are little higher (1.67-2.0eV) than dense $VO_2$ films (**Figure S8**),[48] which could lead to a significant decrease in absorption of short-wavelength range, including visible range, resulting in an enhancement of luminous transmittance ($T_{lum}$). And the blue shift of the absorption edge in the transmission spectra may contribute the main cause in broadening band gap.[49, 50] **Figure 6** shows the comparison of this work with recently reported $VO_2$-based films, indicating that our work in prepared nano-porous $VO_2$ film has an prominent optical performance, more excellent than most of recent reports and our previous works in $VO_2$-based films, including those works fabricated by conventional magnetron sputtering and chemical methods,[1, 3, 17-20, 25, 47, 51-61] even better than some periodic structures those prepared by lithography.[24, 61, 62] Furthermore, this work is in good agreement with the trend for development direction of $VO_2$-based thermochromic films.

The phenomenon of LSPR on $VO_2$ is drawn attention due to its temperature-dependent behavior, which can also affirm the existence of itself and was not observed on pure noble metals.[62-64] The transmittance spectra of nano-porous $VO_2$ film (**Figure 3a**) suggest that LSPR can be observed at 90°C but not at 20°C, which may due to the metallic phase formation above the transition temperature, as well as the existence of ~100nm scale particles, agreed with our previous work.[46] In order to investigate the variation of LSPR peaks position, temperature-dependent spectra were measured from 25°C to 100°C (**Figure S9a**). And the corresponding switched temperature-dependent extinction was calculated based on formula $A=-lg\,(Transmittance)$, which was illustrated in **Figure 3b**. It is evident that LSPR appears at high temperature and shows gradually attenuation with the reduced temperature, especially in the range of 60-80°C (**Figure S8c** and **Figure S8f**). The high-temperature strengthened LSPR phenomenon may ascribe to the emergence of intermediate states during the structural and electronic phase transition in $VO_2$ thin films,[65, 66] which present a temperature-dependent fraction of metallic phase, giving rise to the gradual enhancement of relevant LSPR intensity. Moreover, thermal hysteresis loop of transmittance was also carried out at 2000nm wavelength in IR region, as displayed in **Figure 3c**. The transition temperature ($T_c$) of films were obtain by the first-ordered differential of thermal hysteresis loop and the Gaussian fit of calculated curves (inset the top right corner of **Figure 3c**. It can be calculated that the transition temperature were 75.1°C ($T_{c\text{-heating}}$) and 56.6°C ($T_{c\text{-cooling}}$) corresponding to heating and cooling routes, while the hysteresis loop width is about 18.5°C. Here, the critical metal to insulator transition (MIT) temperature ($T_c$) was defined as the central point of the hysteresis loop $T_c=(T_{c\text{-}h}+T_{c\text{-}c})/2$.[59, 67] Therefore, $T_c$ of nano-porous $VO_2$ film is around 65.9°C, which is similar to the continuous dense films (**Figure S7**) that we prepared with equal thickness. $T_c$ of all the samples are around the MIT critical temperature of bulk $VO_2$ ~68°C.[4] Notably, as for hysteresis loop width, nano-porous $VO_2$ film is wider than dense $VO_2$ film in equal thickness (**Figure S7c** and **Figure S7f**). We speculate that the large size (over 100nm) of $VO_2$ particles could lead to broaden hysteresis loops ($\Delta T_c$).[68] In addition, the void air space between $VO_2$ particles can also impede the heat conduction, resulting in increasing hysteresis both in heating and cooling processes, and broadening width of thermal hysteresis loops.

As is well-known, the dark-brown (orange-yellow) color of $VO_2$ is one of the drawbacks in practical use. Thus, we calculated the CIE spectra of nano-porous $VO_2$ film and the other corresponding dense films based on transmittance spectra, as displayed in **Figure 3e**. It is obvious that the fabricated nano-porous VO$_2$ film (VP) exhibits a light green-yellow color, while the continuous dense films present the orange-yellow color. And the light green-yellow color may be more acceptable than orange-yellow color in actual application because green is the most sensitive color for human eyes. Furthermore, optical photo (**Figure 3f**) presents the color and ultrahigh luminous transmittance intuitively. We can see the leaves through the film clearly, and the VP film exhibits a well color reproduction for objects on the other side. In contrast, the dense film is under-performing (**Figure S9b**).

### 3.4 Models and Theory

To further investigate the principle in optical performance enhancement, mechanism analysis were crucial. To our knowledge, the Maxwell-Garnett theory[69] is suitable for a type of topology, a continuous matrix which is dispersed with nanoparticles.[70] Therefore, inspired by this theory, we assume that a certain amount of voids (or VO$_2$ particles) disperse homogeneously in continuous VO$_2$ films (or continuous air films) with relevant ideal structures: random oriented prolate nanoparticles (**Figure S10**) with the same axial ratio. Then, the optical properties of the "valid" medium consisting of nanoparticles and embedding matrix are dominated by an effective dielectric function $\varepsilon^{MG}$ as follows.

$$\varepsilon^{MG} = \varepsilon_m \frac{1 + \frac{2}{3} f \alpha}{1 - \frac{1}{3} f \alpha}$$

(1)

$$\alpha = \frac{\varepsilon_p - \varepsilon_m}{\varepsilon_m + L(\varepsilon_p - \varepsilon_m)}$$

(2)

Where $\varepsilon_m$ and $\varepsilon_p$ are the dielectric permeability and dielectric function of for matrix and particles respectively, and $f$ is the "filling factor", which can represent the volume fraction occupied by the particles in previous reports.[71, 72] In formula (2), $L$ is the advisable depolarization factor. For spheres, $L=1/3$. And for prolate spheroids and oblate spheroids, depolarization factor had calculated in previous report.[73] In our modeling, to simplify, we assume that all of the particles (or void) are ideal spheres, hence, these depolarization factors were used in following computation. Spectral optical constants ($n, k$) of VO$_2$ film, which is crucial data in optical computation, were measured and fitted by spectroscopic ellipsometer (**Figure S11b**) and had also been used in our previous works.[3, 74] The interrelation between optical constants and dielectric function are following the formulas：

$$\varepsilon_f = \varepsilon_{f1} + i\varepsilon_{f2} \quad (\varepsilon_f = n^2 - k^2, \varepsilon_{f2} = 2nk) \tag{3}$$

In calculation, the magnitude of $\varepsilon_p$ is probably equal to the experimentally determined $\varepsilon_f$ for films but may need correction because the conduction electrons undergo surface scattering in the particles, recent reports had already confirmed that VO$_2$ has an exceptionally small scattering time. Thus, it is appropriate to put $\varepsilon_f \approx \varepsilon_p$ in approximation. Finally, all of the calculations of transmittance were executed from Fresnel's equations[75] and utilize the measured optical constants in advance or the calculated optical constants based on ideal structures above.

Herein, on the basis of Maxwell-Garnett theory, we propose some ideal models, as illustrated in **Figure 4a**. In consideration of the porosity, the integral volume of air (void) is set in value of 0.35 according to the prepared nano-porous VO$_2$ film. **Model 1** is the film which composed of continuous VO$_2$ embedded with spherical voids (pitaya-like structure), while **Model 2** denotes the film formed with VO$_2$ particles dispersed in the air. For **Model 1**, the simulated spectra is presented in **Figure 4b**, which match well with the void film we deposited. Then, for **Model 2**, the simulated spectra exhibit a prominent LSPR absorption peak in metallic states, implying a larger solar modulation ability, as illustrated in **Figure 4c**. However, it is hardly to obtained proper spectra similar to our prepared nano-porous VO$_2$ film by combining these two models. Thus, we provide another facile pattern which is made of continuous VO$_2$ film with perforative pores in vertical direction, as illustrated in **Model 3**. These perforative pores may improve transmittance in full wavelengths, while maintain the continuity of film. Ultimately, **Model 4** is proposed here, which is mixed with the previous three models. The simulated spectra (**Figure 4d**) of **Model 4** is generally consistent with the fabricated nano-porous VO$_2$ film, and the percentage composition is 25%, 40%, 35% for **Model 1-3** in sequence. Furthermore, the calculated refractive indexes (*n*) (**Figure S11b**) express a significant reduction in visible range, which can also elucidate the prominent enhancement in luminous transmittance. For the former three models, void film can enhance the luminous transmittance and solar modulation ability, but cannot enhance a lot. VO$_2$ film with particles brings in LSPR absorption effect, which can largely enhance solar modulation ability but the adhesion between substrates and particles is too weak. VO$_2$ film with perforative pores can improve transmittance in all wavelength and has well adhesion with substrates, but cannot promote the solar modulation ability. Fortunately, this kind of nano-porous film has gathered the advantages of former three models, which can exhibit ultrahigh luminous transmittance accompany with excellent solar modulation ability, and simultaneously maintain a relevant strong adhesion to substrates. In summary, the optical properties of models were collected and listed in **Table 2.** One can see that the optical performance of models are in substantial agreement with our experiments (**Table 1**). Indeed, **Model 4** has a bit larger solar modulation ability than fabricated films due to the stronger LSPR absorption, which may ascribe to the idealization in models and the relatively dispersive particle sizes in prepared films.

**3.5 Energy-Saving Application**

Films suitable for actual use in architectural glazing should possess visible transmittance preferably in excess of 50~60% (as is shown in **Figure 5** with blue dotted line).[6, 23] Therefore, most of the reported VO$_2$ thin films cannot meet the requirement of smart building window (**Figure 5**), while some of the films had unsatisfactory solar modulation efficiency even if they reached a higher visible transmittance. Fortunately, in this work, optical performance satisfied the requirement of smart window in architecture whether for prepared nano-porous VO$_2$ film or double-layer films. Moreover, these films also provide an optimal solar modulation efficiency while maintaining ultrahigh luminous transmittance. It is notable that the luminous transmittance of nano-porous VO$_2$ film is far beyond 60%, and even over 70%, which is the essential requirement of front windshield (red dotted line in **Figure 5**) for automobile according to *Chinese technical specifications for safety of power-driven vehicles operating on roads (GB 7258-2017)* and other previous works.[76, 77] In addition, for energy-saving application, when considering the appropriate *ΔT$_{sol}$* (over 10%), optical performance which matches well with front windshield should be in the yellow area in the top right corner of **Figure 6**. Apparently, rare previous studies could satisfy this condition except this work. In other words, this kind of nano-porous VO$_2$ films would have enormous potential in the front windshield of vehicles, as illustrated in **Figure 6a**. In this case, with the thermochromic film, the front windshield could block lots of infrared radiation from solar energy and stem the inner temperature rise of automobile when it is hot outside, resulting in reducing the energy consumption of air-conditioner. When it is cool outside, windshield could let more infrared radiation enter into the automobile, and keep it warm. Angle-dependent luminous transmittance is an important indicator for windshield, as illustrated in **Figure 6b**. When eye-observation visual angle of driver varies from zero to high-angle gradually, the transmittance of visible light may decrease due to the extended path in dense films. For comparison, Angle-dependent transmittance in visible light were measured for nano-porous $VO_2$ film (**Figure 6c**) and dense $VO_2$ film (**Figure 6d** and **Figure S9c**). It can be found that nano-porous $VO_2$ film mainly retain an ultrahigh visible transmittance when incident angle varying from -60° to +60°. Specifically, transmittance at 550nm (**Figure 6e** and **Figure S9d**) were selected to investigate because it is the most sensitive wavelength for human eyes. One can see that the visible transmittance is nearly over 60% for all incident angles, guaranteeing a highly stable visible transmittance when looking through the windshield in varied eye-observation visual angles (**Figure 6b**), ensuring the transportation safety. Moreover, the optical photo also affirms this point (**Figure S7e**), in comparison with dense $VO_2$ film.

## 4 Conclusion

In this work, a novel homogeneous nano-porous $VO_2$ film was deliberately fabricated by combining magnetron sputtering and specific post annealing process together. Particularly, PTFE acted as a vital pore-forming template in fabrication. As expected, this kind of nano-porous $VO_2$ film exhibit an excellent optical performance with ultrahigh luminous transmittance ($T_{lum-L}$=78.0%) and a prominent solar modulation efficiency ($\Delta T_{sol}$=14.1%). More importantly, nano-porous $VO_2$ film was composed of connected teeth-like nano-particles and nano-pores, giving rise to the enhanced absorption of Localized Surface Plasmon Resonance (LSPR) in near-infrared range (1050-1200nm wavelengths) and increased transmittance in visible light, resulting in optimizing the luminous transmittance and solar modulation ability. In addition, based on Maxwell-Garnett theory, an ideal model was first proposed to simulate and verify the prepared structures, which combines models of void film, film with particles, and continuous film with perforative pores together. And the modeling is in good agreement with the experiment. Moreover, the ultrahigh luminous transmittance of nano-porous film has reached the requirement of windshield, opening a new application field for $VO_2$ thin films, and the facile and universal fabrication process may promote the development of modified nano-structural thin films.

## Associated content

*Supporting Information*

Additional materials and characterizations. Structural characterization including SEM, XRD, AFM images and Raman shift spectra for $VO_2$-based films; The SEM images and formation process of collapsed films; Transmittance spectra of dense films and double-layer VP films; Thermal hysteresis loops; Calculation of optical band gaps; Eye-observation virtual angle measurements; Idea model based on Maxwell-Garnett theory; Calculated optical constants for $VO_2$ films.

## Acknowledgment

This study was financially supported by the National Natural Science Foundation of China (NSFC, No.51572284), the Youth Innovation Promotion Association, Chinese Academy of Sciences (No.2018288), the Science Foundation for Youth Scholar of State Key Laboratory of High Performance Ceramics and Superfine Microstructures (No. SKL201703), the Shanghai Pujiang

2018, **120**, 186103.

**Figures**

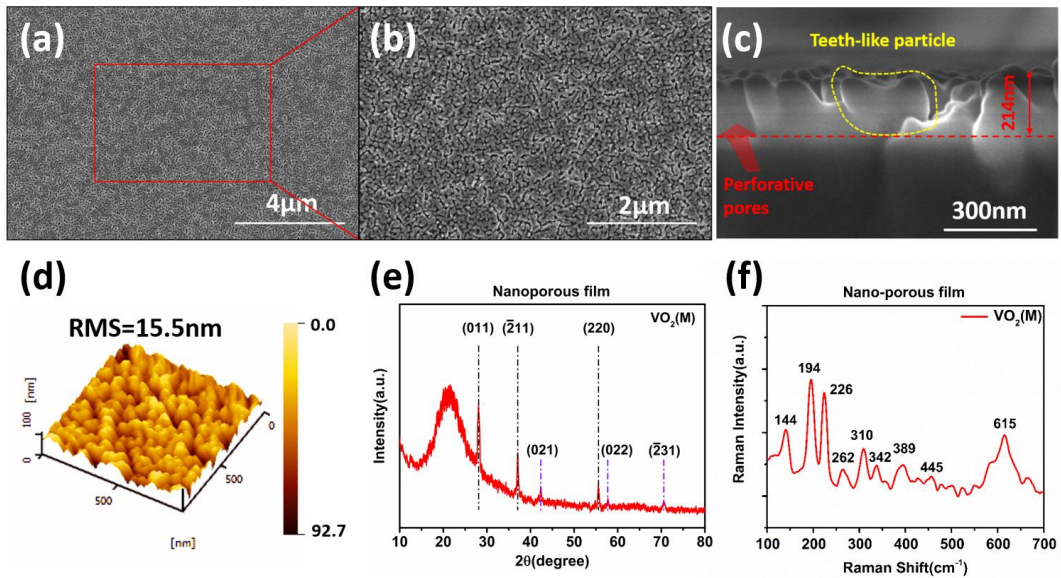

**Figure 1.** (a) Large-area FESEM images of nano-porous VO$_2$ film growing by sputtering and post annealing process and the corresponding (b) amplified surficial image. (c) Cross-section FESEM image of nano-porous VO$_2$ film. (d) AFM image of surficial morphology for nano-porous VO$_2$ film and the corresponding Root-Mean-square (RMS) roughness. (e) XRD pattern of prepared nano-porous VO$_2$ film. (f) Raman spectra of the nano-porous VO$_2$ film.

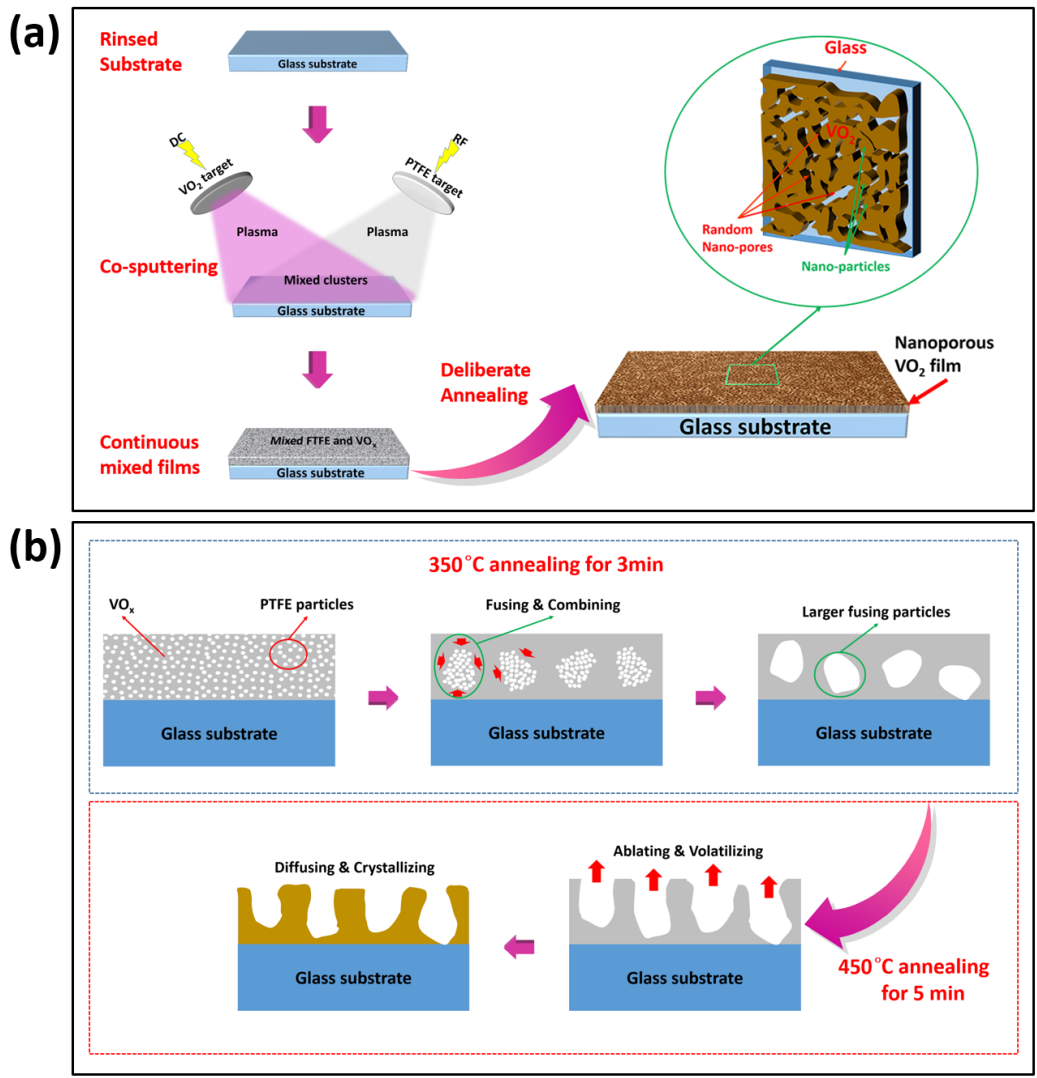

**Figure 2.** (a) Schematic of fabrication route for nano-porous VO$_2$ films. (b) Mechanism in form of nano-pores films at post annealing process.

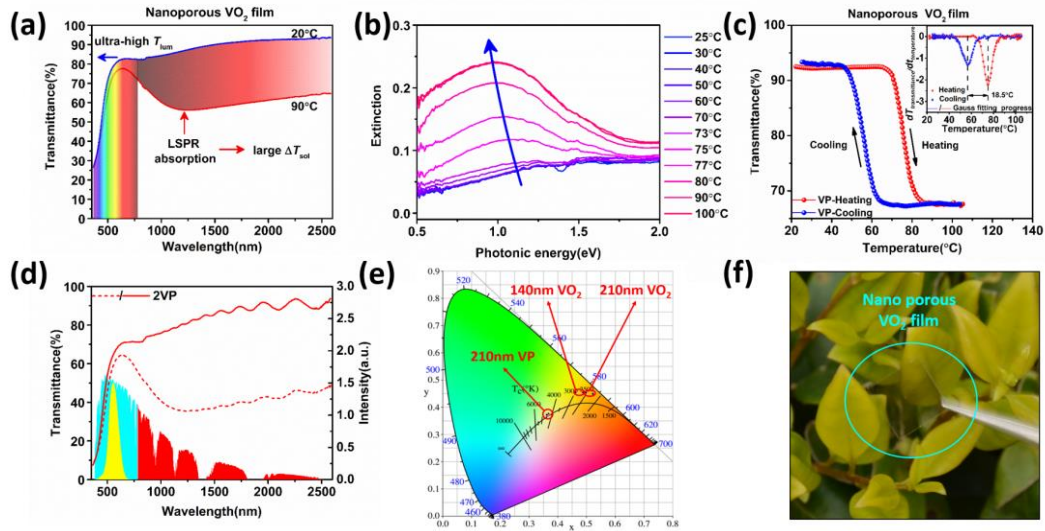

**Figure 3.** (a)Transmittance spectra of the nano-pores VO$_2$ films. (b) Temperature-dependent LSPR peak variation of the nano-pores VO$_2$ films in temperature range from 25 to 100 °C. Extinction (A) is calculated *via A=-lg (transmittance)*. (c) Temperature-dependent hysteresis loops and the corresponding first-order differential curves inset the top right corner. (d) Transmittance spectra of the double-layer nano-pores VO$_2$ films. The coloring areas indicate the normalized values of spectral irradiance corresponding to the visible (cyan) and NIR (red) ranges of solar spectra, and the yellow area indicates the values of eye sensitivity function. (e) CIE image of nano-pores VO$_2$ film compared with continuous VO$_2$ films. (f) Optical photo of the nano-pores VO$_2$ film on quartz.

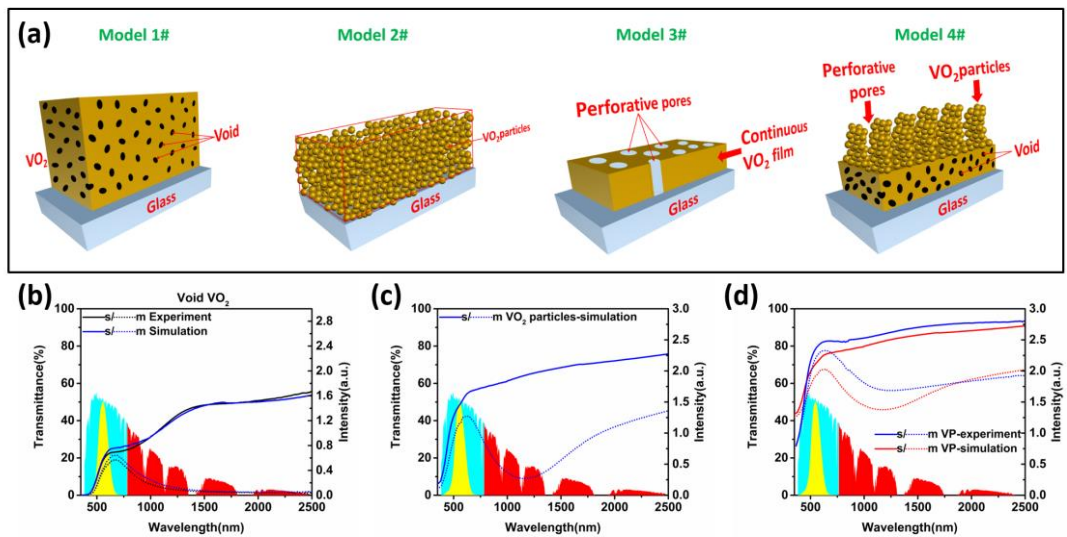

**Figure 4.** (a) Different film models based on Maxwell-Garnett theory, including film with void (Model 1#), film with particles (Model 2#), film with perforative pores (Model 3#) and the mixed model of three formers (Model 4#). Experimental and simulated spectra for (b) Model 1# and (c) Model 2#. (d) Spectra of the fabricated nano-porous VO$_2$ film and the simulated spectra for mixed Model 4#, and the corresponding wavelength-dependent solar energy distribution.

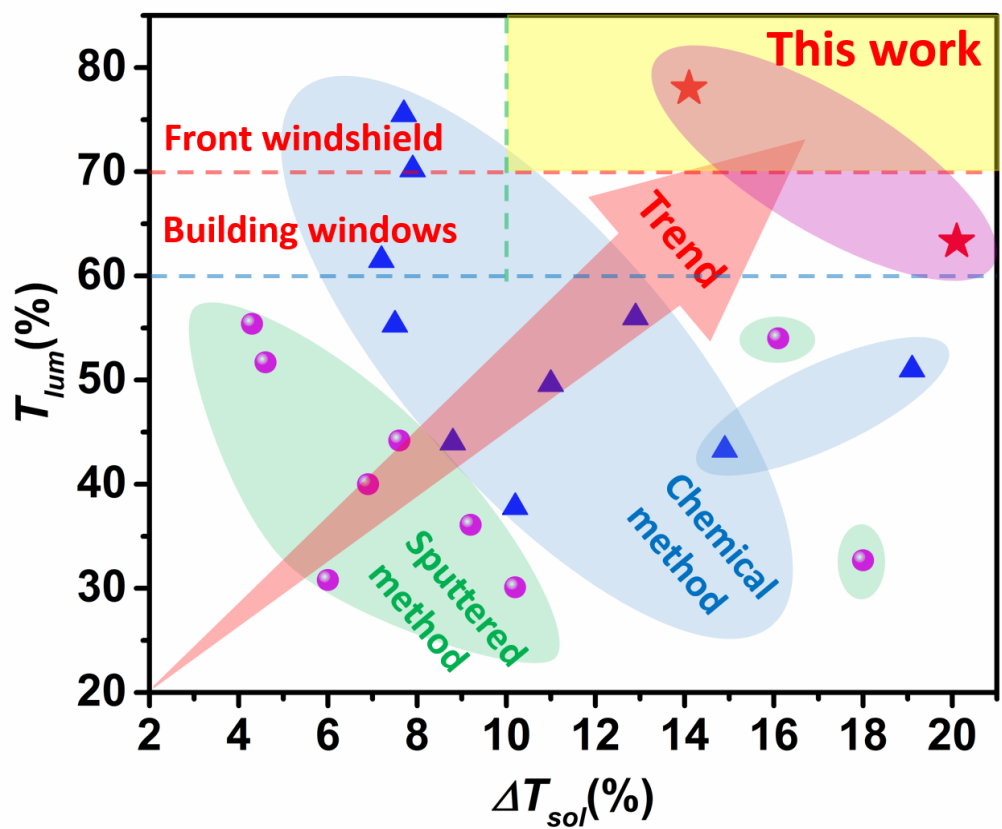

**Figure 5.** Comparison of this work with recently reported $VO_2$-based thermochromic films prepared by different methods.

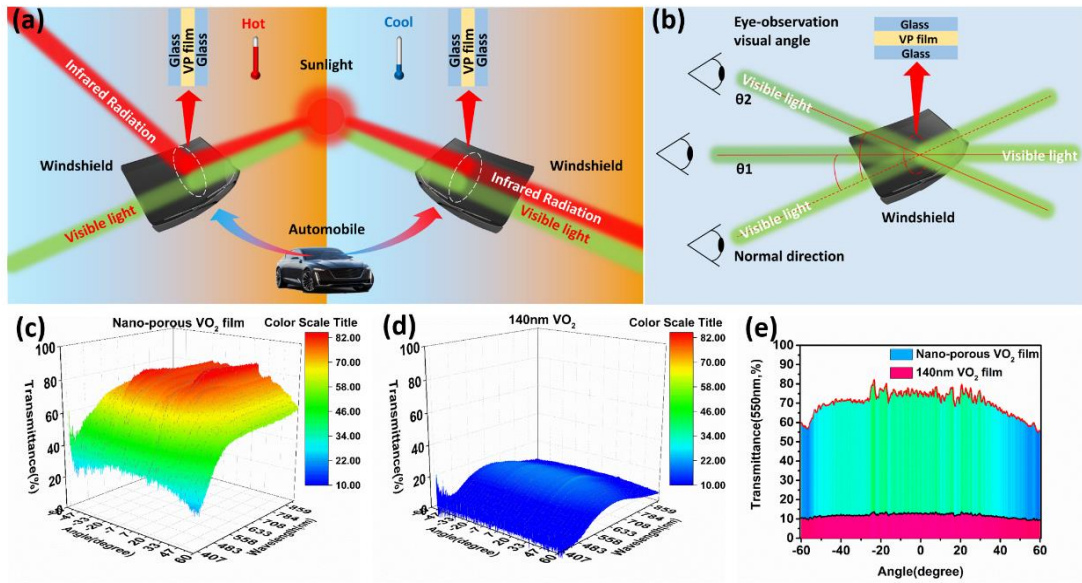

**Figure 6.** (a) Schematic diagram of front windshield combined with nano-porous VO$_2$ film in automobile. (b) Schematic diagram of different eye-observation visual angles through front windshield with nano-porous VO$_2$ film. Three-dimensional angle-dependent (-60°~+60°) luminous transmittance for (c) nano-porous VO$_2$ film and (d) dense VO$_2$ film. (e) angle-dependent (-60°~+60°) transmittance at extremum of the eye sensitivity function (550nm) for nano-porous VO$_2$ film and dense VO$_2$ film.

**Tables**

**Table 1. Optical properties for prepared films.**

| Sample | $\Delta T_{sol}$ | $T_{lum-L}$ | $T_{lum-H}$ |
|---|---|---|---|
| 50nm VO$_2$ | 10.0 | 39.3 | 35.2 |
| 140nm VO$_2$ | 15.8 | 17.2 | 12.2 |
| 210nm VO$_2$ | 14.8 | 9.8 | 6.4 |
| Nano-porous VO$_2$ film (VP) | 14.1 | 78.0 | 72.8 |
| 2VP | 20.1 | 63.3 | 57.1 |

**Table 2. Optical properties for simulated models.**

| Sample | $\Delta T_{sol}$ | $T_{lum-L}$ | $T_{lum-H}$ |
|---|---|---|---|
| Void VO$_2$ film | 15.0 | 18.2 | 12.8 |
| VO$_2$ particles film | 26.9 | 47.8 | 37.3 |
| Nano-porous VO$_2$ film (VP) | 17.7 | 70.7 | 63.9 |

# Supporting Information:

# Sputtered Spontaneously Nano-porous VO$_2$-based Films *via* PTFE Self-Template: Localized Surface Plasmon Resonance Induced Robust Optical Performance for Solar Glazing Application

Shiwei Long, [a,b,c] Xun Cao, [a,b,]* Rong Huang, [d] Fang Xu, [a,b,c] Ning Li, [e] Aibin Huang, [a,b] Guangyao Sun, [a,b] Shanhu Bao, [a,b] Hongjie Luo [f] and Ping Jin [a,b,g,]*

[a] State Key Laboratory of High Performance Ceramics and Superfine Microstructure, Shanghai institute of Ceramics, Chinese Academy of Sciences, Shanghai, 200050, China.
[b] Research Center for Industrial Ceramics, Shanghai Institute of Ceramics, Chinese Academy of Sciences, Shanghai 200050, China
[c] University of Chinese Academy of Sciences, Beijing 100049, China
[d] Key Laboratory of Polar Materials and Devices, Ministry of Education, East China Normal University, Shanghai 200241, China
[e] Department of Materials Science and Engineering, College of science, China University of Petroleum Beijing, No. 18 Fuxue RD, Beijing 102249, China
[f] School of Materials Science and Engineering, Shanghai University, Shangda Road. 99, Baoshan, Shanghai 200444, China
[g] Materials Research Institute for Sustainable Development, National Institute of Advanced Industrial Science and Technology, Nagoya 463-8560, Japan

**Corresponding authors**
State Key Laboratory of High Performance Ceramics and Superfine Microstructure, Shanghai Institute of Ceramics, Chinese Academy of Sciences, Shanghai 200050, China
*E-mail：cxun@mail.sic.ac.cn (X. Cao)
*E-mail：p-jin@mail.sic.ac.cn (P. Jin)

## Figures
*Structural characterization.*

The void ratio of the nano-porous film was estimated by computing the surficial area ratio of pores at SEM image (**Figure S1b**) using *ImageJ* software, and the sizes of pores were also counted by the software, as shown in **Figure S1c-d**. The emulated void ratio is around 35%-40%, and the sizes of pores is about 50-200nm. In addition, for comparison, XRD patterns of dense films were measured (**Figure S2a**), demonstrating the pure phase of $VO_2$ (M). The deeps of some pores have also been measured by AFM (**Figure S2b**), which is about 70nm. Moreover, Raman shift spectra of 50nm and 140nm dense films were illustrated in **Figure S3**, which could also confirm the formation of $VO_2$.[1, 2]

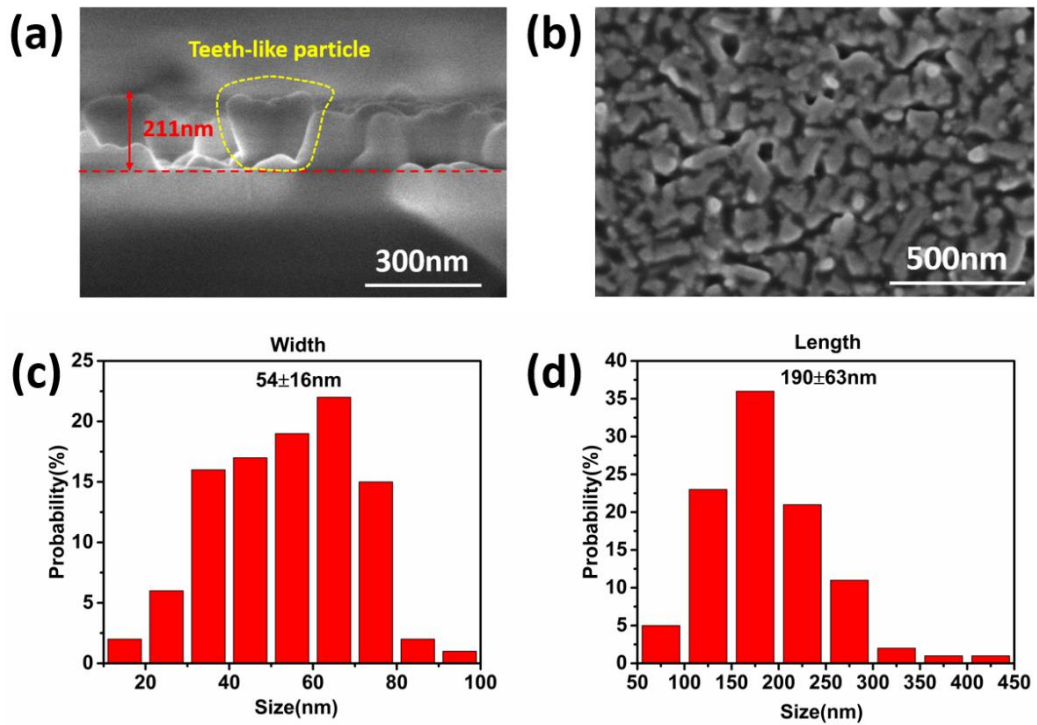

**Figure S1.** (a) Cross-section and (b) surface FE-SEM images of nano-porous $VO_2$ films. (c) Width and (d) length distribution of nano-pores.

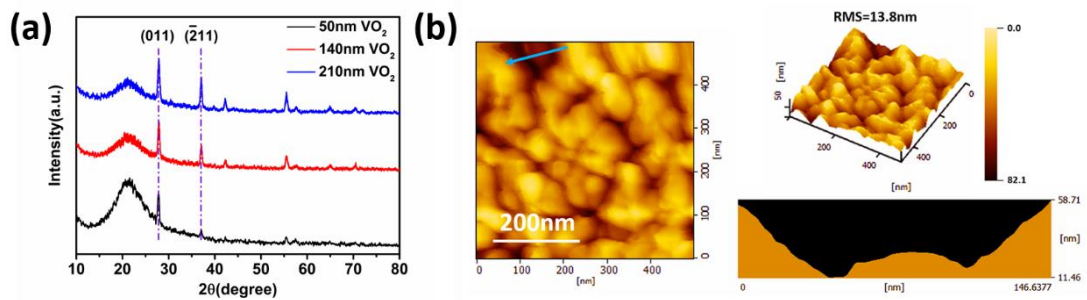

**Figure S2.** (a) XRD patterns of dense films. (b) AFM image of surficial morphology and the depth of the pores.

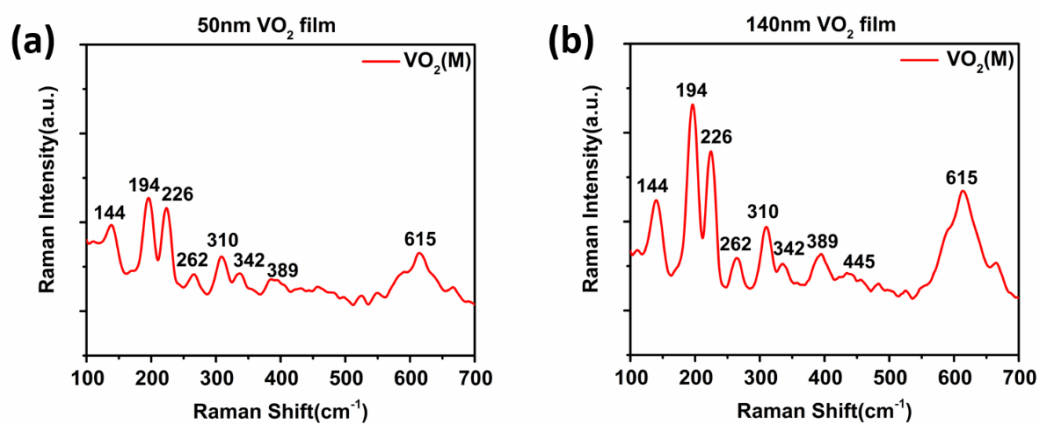

**Figure S3.** Raman shift spectra of the dense (a) 50 nm $VO_2$ film and (b) 140 nm $VO_2$ films.

*Collapsed films.*

  The prepared nano-porous film has a larger pore size than annealed dense films (**Figure S4a-b**). In the post annealing route, when we use one-step annealing, maintain the temperature at 450°C for 8min, the surface of the film displayed some cracks and collapsed into pieces (**Figure S4c**), and the film is heterogeneous and has an inferior adhesion on the substrate. Meanwhile, the pores are non-uniform and even reached microscale range (**Figure S4d-e**), and the film exhibited a poor crystallinity (**Figure S4f**). We speculate that a violent ablating and volatilizing were happened in annealing process (**Figure S5**) when was lack of temperature buffer, resulting in collapsing of film.

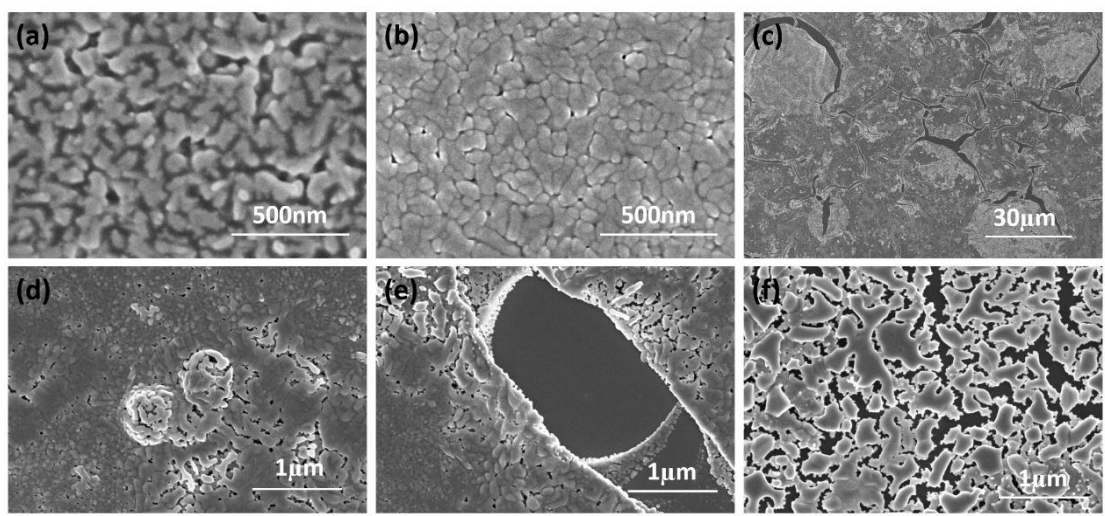

**Figure S4.** Surficial FE-SEM images of prepared (a) nano-porous film and (b) annealed dense film. Surficial FE-SEM images of (c), (d), (e), (f) prepared films treated in one-step annealing at 450°C for 8min.

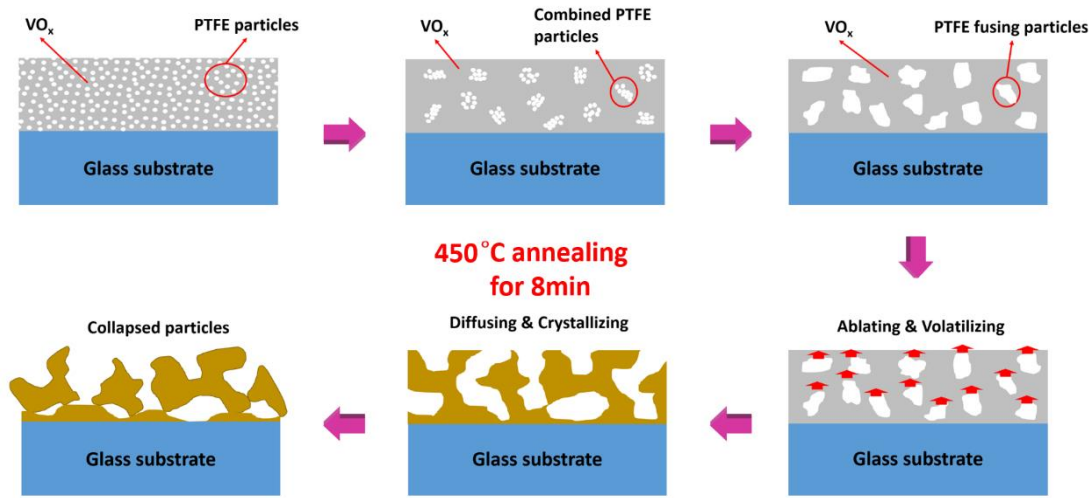

**Figure S5.** Schematic of the formation of the collapsed $VO_2$ film in one-step annealing process.

*Optical performance.*

Optical performance of 50 nm dense $VO_2$ film and equal thickness dense film had been measured in comparison with our nano-porous $VO_2$ film (VP), as exhibited in **Figure S6a-b**. It is obvious that the luminous transmittance of VP is much higher than 50 nm dense $VO_2$ film and the equal thickness dense film. In addition, the schematic of double-layer VP films was designed in **Figure S6c**.

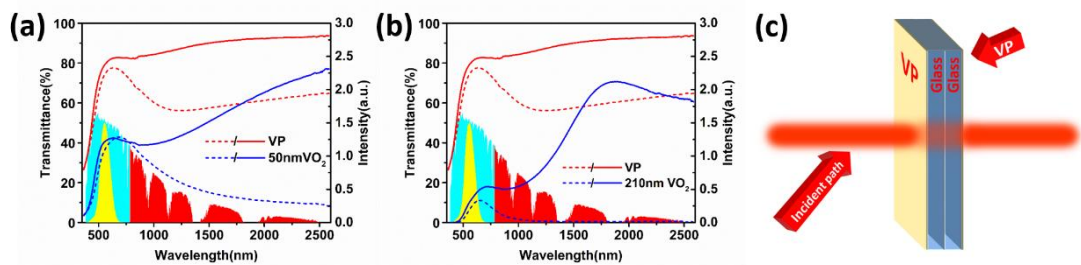

**Figure S6.** Transmittance spectra of deposited (a) 50nm and (b) 210nm dense $VO_2$ films. (c) Schematic of double-layer VP films.

*Hysteresis loops.*

Thermal hysteresis loops of transmittance were also carried out at 2000nm wavelength in heating/cooling processes for continuous dense films. The transition temperature of films were obtain by the first-ordered differential of thermal hysteresis loop and the Gaussian fit of calculated curves. As is shown in **Figure S7**, the $T_c$ are 64.4°C, 64.4°C and 62.4°C for 50nm, 140nm and 210nm VO2 films respectively, while the corresponding widths of their hysteresis loops are 12.3, 14.4 and 16.6°C. Their $T_c$ are nearly the critical transition temperature of bulk $VO_2$ (~68°C).[3] The widths of hysteresis loops were broadened as the thickness increased. Particularly, the hysteresis loop width of VP film is wider than all of above.

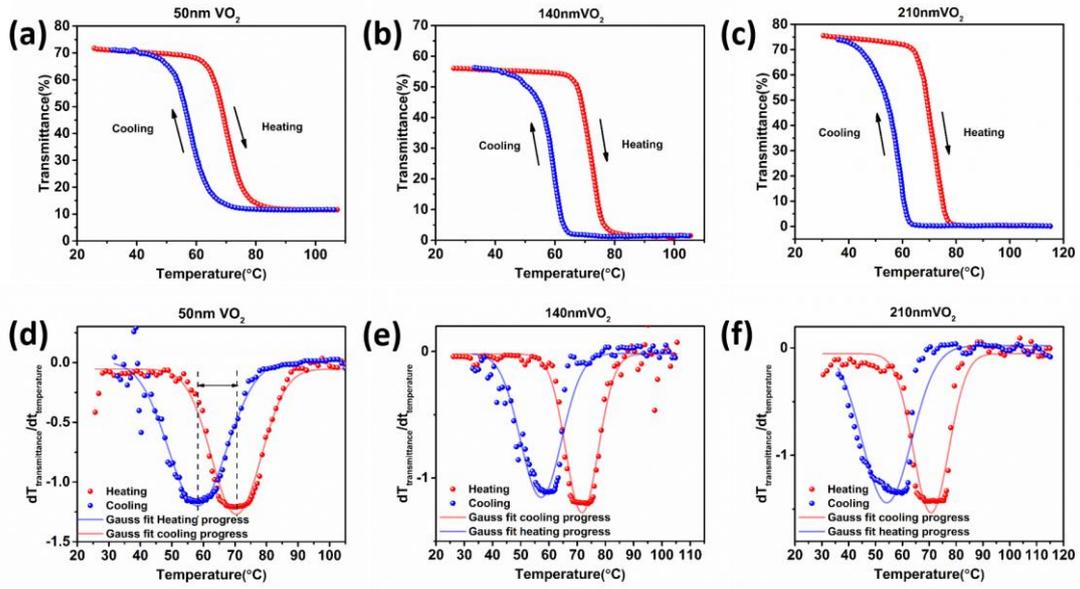

**Figure S7.** Thermal hysteresis loops for prepared (a) 50nm, (b) 140nm and (c) 210nm $VO_2$ films, and the corresponding first-order differential curves of hysteresis loops for (d) 50nm, (e) 140nm and (f) 210nm $VO_2$ films respectively.

*Optical energy band gap.*

Optical band gap ($E_g$) of all the samples are derived from the expression[4]:
$$(\alpha h\nu)^{1/2} = A(h\nu - E_g)$$

Where α is the absorption coefficient, A is a constant and *h* is the photon energy. By plotting *(α)$^{1/2}$* vs *hν*, it can be linear fitted, and the $E_g$ is obtained by the intercept of the fitted line and the hν axis (*α=0*).

Based on the transmittance spectra and temperature-dependent spectra (**Figure S9a**) of films, the optical band gap (**Figure S8a-b, d-e**) and temperature-dependent LSPR peaks (**Figure S8c, f**) are obtained. The band gaps were 2.0, 1.55, 1.86 and 1.73eV for VP, 50V, 140V and 210V respectively in semiconductor states, while were 1.67, 1.74, 1.64 and 1.66eV in metallic states. One can see that the band gap of nano-porous film (VP) is slight lager than dense film, which is around 1.67-2.0eV.

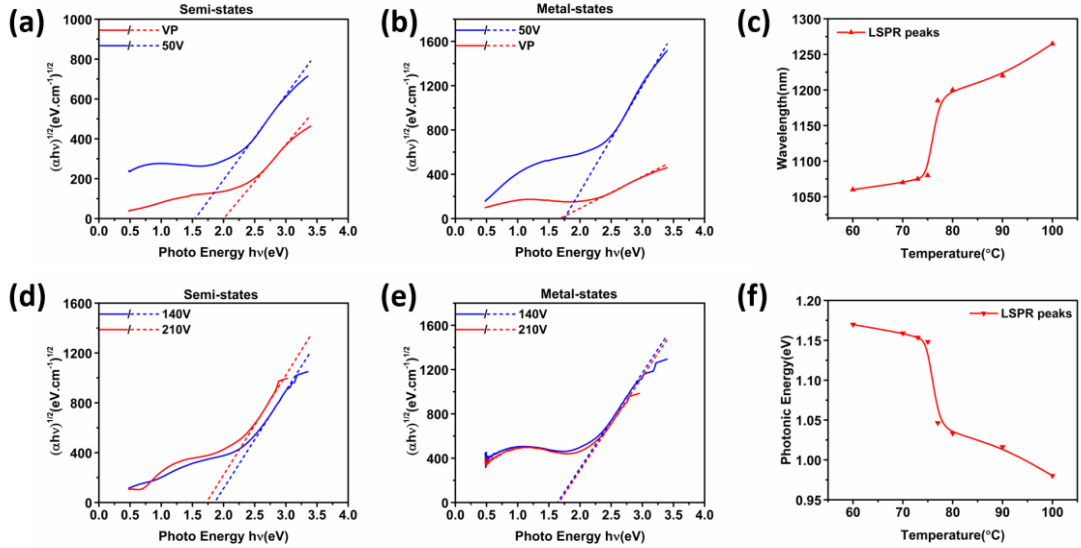

**Figure S8.** $(\alpha h\nu)^{1/2}$ and plots for all the prepared $VO_2$ films at (a), (d) semiconductor states and (b),(e) metallic states. Temperature-dependent variation of LSPR peaks changing in (c) wavelength and (f) photonic energy.

*Eye-observation virtual angle.*

Optical photo of dense $VO_2$ film was also obtained (**Figure S9b**). The color of leaves through the film is obviously darker than the realistic color, due to the lower luminous transmittance. Furthermore, Eye-observation from different virtual angles has been photographed in **Figure S9e.** It can been easily to discover that eye-observation from different virtual angles (90, 60 and 30°) is clearly to see the letters behind VP film. On the contrast, it seems hard to observe the letters from varied angles through dense film, especially from a low angle.

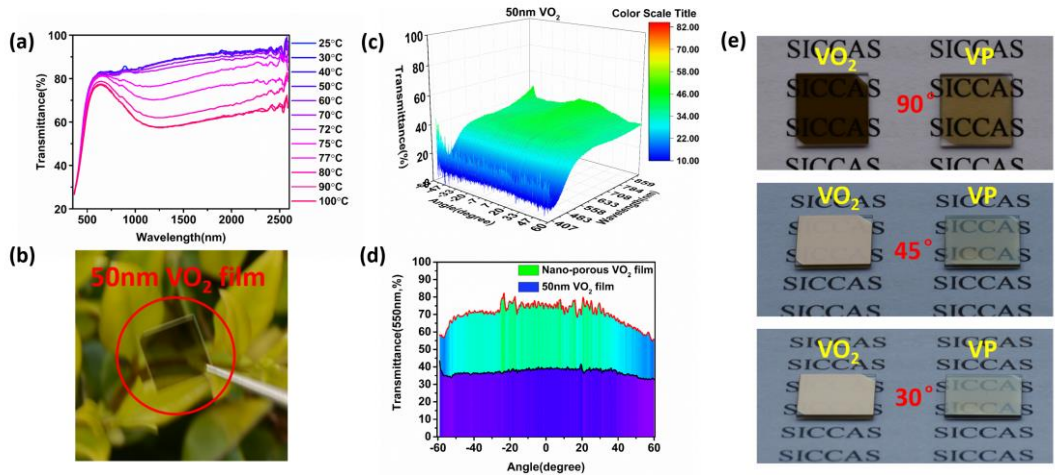

**Figure S9.** (a)Temperature-dependent spectra of VP film. (b) Optical photo of the dense $VO_2$ film. (c) Three-dimensional angle-dependent (-60°~+60°) luminous transmittance for (c) nano-porous $VO_2$ film and 50nm dense $VO_2$ film. (d) angle-dependent (-60°~+60°) transmittance at extremum of the eye sensitivity function (550nm) for nano-porous $VO_2$ film and 50nm dense $VO_2$ film. (e) Optical photos from different observation angles for dense $VO_2$ and VP films.

*Models and theory.*

Maxwell-Garnett theory[5] is appropriate for a topology with nanoparticles dispersed in a continuous matrix[6]. Thus, we set up the models based on this theory. Here, we assume that a certain amount of voids (or $VO_2$ particles) disperse homogeneously in continuous $VO_2$ films (or continuous air) with pertinent ideal structures[7]: random oriented prolate nanoparticles with the same axial ratio (**Figure S10**). In the modeling, optical constants (n, k) were required to simulate the spectra. Hence, we obtained the simulation by using the optical constants derived from the previous deposited dense $VO_2$ film (**Figure S11b**) and the optical constants of void film and film with particles were calculated and shown in **Figure S11a-b**. One can see that the value of refractive index has decreased a lot for both of particle's film and void film in visible range, leading to a reduce in refraction and the luminous transmission has been enhanced.

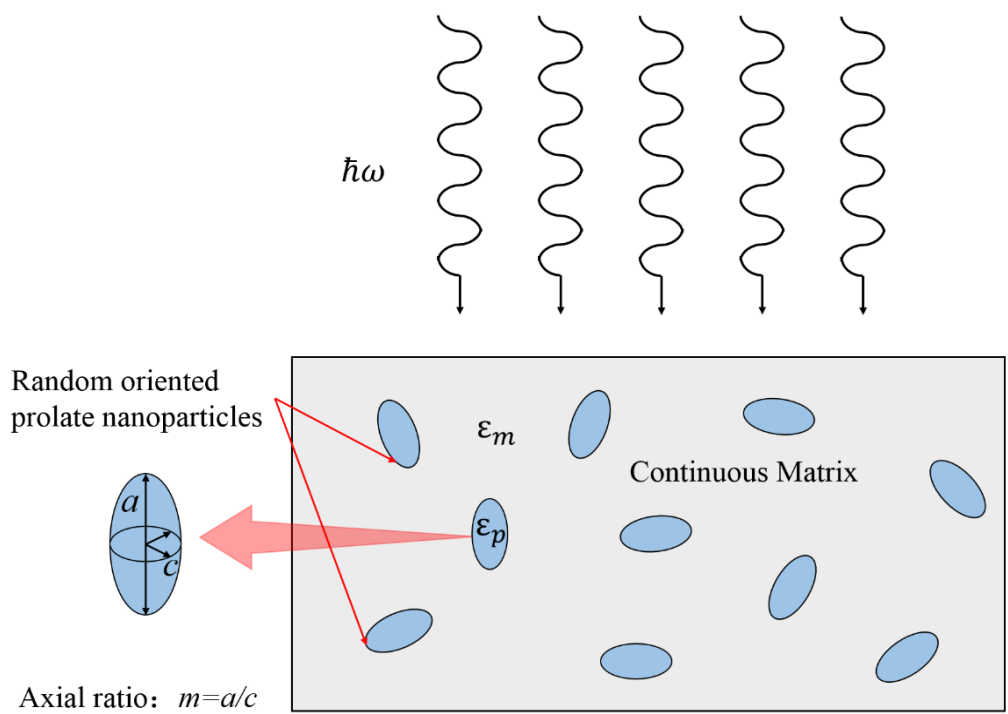

**Figure S10.** Schematic diagram that the ideal random oriented prolate nanoparticles disperse homogeneously in continuous matrix.

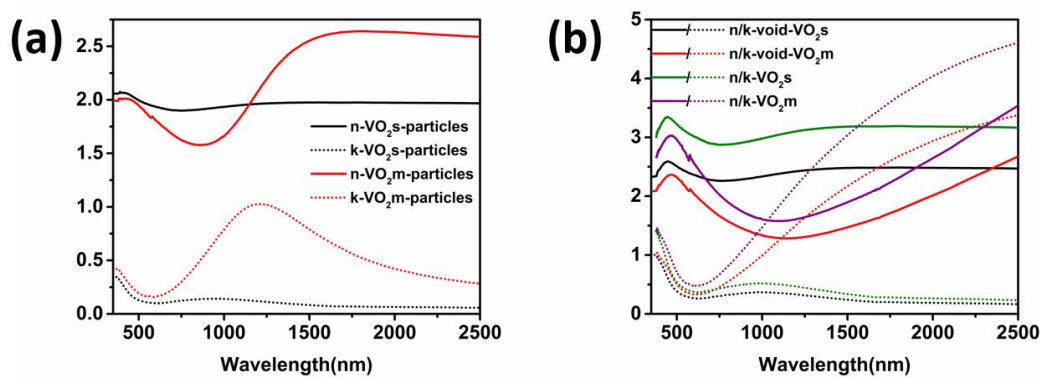

**Figure S11.** Optical constants (n, k) for (a) $VO_2$ film formed of particles as well as (b) void $VO_2$

film and dense $VO_2$ film.